\documentclass[12pt]{iopart}

\usepackage{graphicx}
\usepackage{amssymb}

\begin{document}

\title{Time-dependent currents of 1D bosons in an optical lattice}

\author{J.~Schachenmayer, G.~Pupillo and A.~J.~Daley}

\address{Institute for Theoretical Physics, University of Innsbruck,
  A-6020 Innsbruck, Austria} 

\address{Institute for Quantum Optics and Quantum Information of the
  Austrian Academy of Sciences, A-6020 Innsbruck, Austria}

\date{\today}

\begin{abstract}
  We analyse the time-dependence of currents in a 1D Bose gas in an
  optical lattice. For a 1D system, the stability of currents induced
  by accelerating the lattice exhibits a broad crossover as a function
  of the magnitude of the acceleration, and the strength of the
  inter-particle interactions. This differs markedly from mean-field
  results in higher dimensions. Using the infinite Time Evolving Block
  Decimation algorithm, we characterise this crossover by making
  quantitative predictions for the time-dependent behaviour of the
  currents and their decay rate. We also compute the time-dependence
  of quasi-condensate fractions which can be measured directly in
  experiments. We compare our results to calculations based on
  phase-slip methods, finding agreement with the scaling as the
  particle density increases, but with significant deviations near
  unit filling.
\end{abstract}


\maketitle

\section{Introduction}

Exciting progress in experiments with cold atoms in optical lattices
\cite{bloch08} has not only paved the way for study of the quantum
phases associated with strongly interacting many-body systems
\cite{jaksch05, lewenstein07}, but also for study of non-equilibrium
dynamics in such systems. This is particularly true for transport
properties, where the long coherence times associated with the
experiments make it possible to gain new insight into phenomena such
as spin-charge separation \cite{kollath05, kollath06, kleine08}, and
currents in the presence of impurities and junctions \cite{micheli04,
  daley05, daley08, tokuno08} by studying them in a new
environment. Particularly in the case of 1D systems, this connection
to non-equilibrium dynamics is further strengthened by the recent
development of numerical methods based on matrix product states
\cite{verstraete08}, including the time evolving block decimation
(TEBD) algorithm \cite{vidal03, vidal04} and the related
time-dependent density matrix renormalisation group (t-DMRG) methods
\cite{daley04, white04}. These make possible the quantitative
prediction of time-dependent dynamics for size scales that are typical
in experiments, as well as the identification of parameter regimes in
which specific phenomena can be observed.

\begin{figure}[htb]
  \centering
 \includegraphics[width=0.7\textwidth]{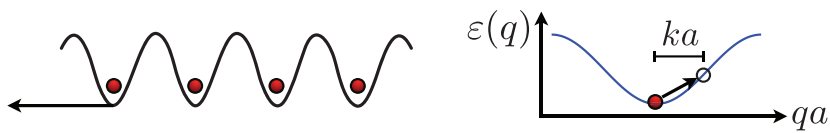}
 \caption{\label{fig:sketch} Bosons in the ground state configuration
    in an optical lattice are instantaneously accelerated to a mean
    quasi-momentum in the lowest Bloch band. The stability of the
    resulting current will depend on the value of this quasi-momentum
    and the inter-particle interactions.}
\end{figure}

A key characteristic for the transport of bosons in a lattice is the
stability of currents in the presence of interactions which can arise
in regimes where the system is superfluid.  However, this stability
depends not only on the current and the interactions \cite{burger01,
  cataliotti03, wu01, smerzi02}, but also on the dimensionality of the
system. This was demonstrated in recent experiments by observing
damping of the centre of mass motion for bosons oscillating in an
harmonic trap \cite{fertig05}. For sufficiently small initial
displacements, the motion was found to be stable up to a critical
lattice depth in higher dimensions, whilst for bosons confined to move
along one dimension, decay of the oscillatory motion was observed for
arbitrarily small depth. This was explained by the increased role of
quantum fluctuations in the 1D system \cite{polkovnikov04,
  gea-banacloche06, pupillo06, rigol05, ruostekoski05, montangero09,
  danshita09}. A similar situation occurs in the case where a
homogeneous current is created, e.g., by starting in the lowest energy
state, and then accelerating the lattice \cite{fallani04, cristiani04,
  sarlo05, mun07, ferris08} (as shown in figure~\ref{fig:sketch}). A
mean-field stability diagram as a function of the magnitude of the
acceleration and the strength of inter-particle interactions was
computed by Altman et al. \cite{altman05, polkovnikov05}, predicting a
sharp transition between stable and unstable regions. For bosons
confined in 3D, such transitions were observed recently
\cite{mun07}. However, the same experiments indicated strong
deviations from this behaviour with a 1D gas, for which a crossover
was observed between the two regimes.

Here we perform a detailed numerical analysis of the decay of currents
for a Bose gas moving uniformly in 1D, making quantitative predictions
for the time-dependence of currents that can be measured in an
experiment. This analysis is made possible via the use of the infinite
TEBD (iTEBD) algorithm \cite{vidal07}, which until now has primarily
been applied to study the ground states of homogeneous systems. Here
it is used to investigate the time-dependence of homogeneous current
flow, allowing predictions to be made over longer timescales without
boundary effects. We begin from the ground state of the Bose-Hubbard
model \cite{jaksch05}, which describes the lattice gas, and consider
an acceleration applied in order to produce a finite current. As a
function of the interaction strength in the gas and the magnitude of
the acceleration, we observe a broad crossover between regions of
stable and unstable current, which we characterise via the variation
in the rate of decay of the current for different parameters. We find
that the crossover region occurs at significantly smaller values of
interaction strength and initial acceleration than are predicted by
mean-field methods. We also compare our decay rates to those obtained
for systems with large filling factors via phase-slip calculations by
Polkovnikov et al.~\cite{polkovnikov05}.

The rest of this paper is organised as follows: In
section~\ref{sec:system} we discuss the Bose-Hubbard model for the
system we study, and the methods we use to characterise the currents
in the system. In section~\ref{sec:numprod} we describe the use of the
iTEBD algorithm and the procedure we use in our calculations.  In
section~\ref{sec:currentresults} we present a quantitative analysis of
the decay of currents within the system, as well as the dependence of
the decay on the initial acceleration and the inter-particle
interactions. The scaling properties of the decay rates are
investigated in section~\ref{sec:phaseslip}, and compared to
calculations based on phase-slip methods, and in
section~\ref{sec:summary} we give a summary and outlook for this
work. \ref{app:blockconv} provides detailed information about
translationally invariant block sizes within the iTEBD algorithm.

\section{Moving 1D bosons in an optical lattice}
\label{sec:system}

In this section we introduce the system of bosons moving in 1D in an
optical lattice. We first introduce the Hamiltonian describing the
system in \ref{sec:bose-hubbard-model}, before discussing the current
and general expectations for its stability during time evolution in
\ref{sec:curr-bose-hubb}. In order to make stronger connections to
quantities that can be measured in an experiment, we introduce the
quasi-condensate fraction in \ref{sec:quasi-cond-fract}, the decay of
which is strongly related to the decay of the current.

\subsection{Bose-Hubbard model}
\label{sec:bose-hubbard-model}

The dynamics of bosons in the lowest band of an optical lattice is
described by the Bose-Hubbard model \cite{bloch08, jaksch05,
  lewenstein07} with Hamiltonian ($\hbar \equiv 1$):
\begin{equation} \label{eqn:bh-h}
  \hat H  
  =  
  -J 
  \sum_{\langle ij\rangle} 
  \hat b^{\dag}_{i} \hat b^{\ }_{j}  
  + 
  \frac{U}{2} 
  \sum_i 
  \hat n_i (\hat n_i - 1) 
  .
\end{equation}
Here, $\hat b_i$ ($\hat b^{\dag}_i$) are the bosonic annihilation
(creation) operators at lattice site $i$, $\hat n_i=\hat b_i^\dag \hat
b_i$, $U$ is the on-site interaction energy shift, and $J$ is the
tunnelling amplitude between neighbouring sites, with $\sum_ {\langle
  ij \rangle}$ denoting a sum over neighbouring lattice sites. This
model is valid in the regime where $U\bar n ,J \ll \omega_T$, where
$\omega_T$ is the energy separation between the lowest two Bloch
bands, and we denote the filling factor of $N$ particles on $M$
lattice sites as $\bar n \equiv N/M$.

For integer $\bar n$, a phase transition is observed in the ground
state of this model as a function of $u\equiv U/J$ between superfluid
and insulating behaviour \cite{fisher89}. In one dimension with $\bar
n = 1$, the critical value of $u$ is $u_c\approx 3.37$
\cite{kuhner00}. The superfluid (SF) ground state is characterised by
quasi off-diagonal long range order in the single particle density
matrix (SPDM) $\langle \hat b_i^{\dag} \hat b^{\ }_j \rangle$. For $u
> u_c$, the system is in a Mott insulator (MI) phase, with
exponentially decaying off-diagonal elements of the SPDM $\langle \hat
b_i^{\dag} \hat b^{\ }_{j} \rangle$. In what follows below, we will
assume that the system is prepared in the ground state of the
Bose-Hubbard model for a particular choice of $\bar n$ and $u$, and
then accelerated to produce a finite initial current.

\subsection{Currents in the Bose-Hubbard model}
\label{sec:curr-bose-hubb}

By accelerating the lattice or applying a linear gradient potential
for a short period of time, it is possible to create a current of
atoms moving with respect to the optical lattice. This can be
quantified via the current operator
\begin{equation} \label{eqn:j-def}
  \hat j_k 
  = 
  \frac{J}{i} 
  \left(
    \hat b^{\dag}_{k+1} \hat b^{\ }_{k}  
    - 
    \hat b^{\dag}_{k} \hat b^{\ }_{k+1} 
  \right) 
  ,
\end{equation}
which appears in the continuity equation
\begin{equation}
  \hat j_k - \hat j_{k-1} 
  \equiv 
  \frac{d}{dt} \hat n_k
  .
\end{equation}
Note that the average of the current expectation values $\sum^M_m
\langle \hat j_m \rangle/M$ is proportional to the mean group velocity
calculated via the quasi-momentum distribution of the particles. In
particular, we can define bosonic operators for the quasi-momentum
modes, $\hat a_q$, which are related to $\hat b_i$ as $\hat b_m \equiv
\sum_{q} e^{-i q a m} \hat a_q /\sqrt{M}$, where $a$ is the lattice
spacing, and $q=2\pi r/L$ for $r\in (-M/2, M/2]$ an integer, with
$L=aM$ the lattice length. We see that
\begin{equation} \label{eqn:average}
  \frac{1}{M} \sum^M_m \langle \hat j_m \rangle 
 =
  \frac{1}{M} \sum_{q} 
  2 J \sin{\left( q a \right)} n_{q} 
  ,
\end{equation}
where $n_{q}\equiv \langle \hat a^{\dag}_{q} \hat a^{\ }_{q} \rangle $
is the quasi-momentum distribution function at quasi-momentum $q$, for
which the corresponding group velocity per lattice constant is given
by $2J\sin(qa)$. In this study, the current will always be
translationally invariant, and so we omit the site label and write
$\langle \hat j \rangle\equiv \langle \hat j_k \rangle$, for any $k$.

If a finite current is induced to the system by accelerating the
lattice so that the quasi-momentum distribution is shifted by an
amount $ka$ (see figure~\ref{fig:sketch}), then we expect the
resulting behaviour to depend on the superfluidity of the gas. Even in
the presence of interactions, a superfluid flow without dissipation
can exist. However, if a current is generated in a non-superfluid
initial state, or if the flow becomes dynamically unstable, then the
current will decay. Whether the current is stable or not is thus a
function of both the initial state of the system and the initial mean
quasi-momentum, $ka$ \cite{polkovnikov05}.

As has been predicted theoretically in the weakly interacting
Gross-Pitaevskii regime \cite{wu01, smerzi02} and demonstrated in
experiments \cite{cataliotti03, fallani04, cristiani04}, the current
will be unstable if the initial superfluid ground state (with all
atoms near $ka=0$) is accelerated to the inverted part of the lowest
Bloch band, i.e., when $ka>\pi/2$ (this corresponds to a classical
instability). We thus expect stable currents only when $ka<\pi/2$.  If
we begin with the ground state of a system with integer filling $\bar
n$, then we also do not expect stable superfluid currents above the
critical interaction $u_c$ for the SF-MI phase transition. A stability
diagram, interpolating between these two regimes was investigated by
Altman et al.~\cite{altman05} using a Gutzwiller mean-field technique,
and they found a sharp transition between stable and unstable
regions. This sharp transition was observed in an experiment by Mun et
al. \cite{mun07} for bosons allowed to move in the lattice in three
dimensions. However, the same experiment observed behaviour more
characteristic of a crossover between stable and unstable currents in
1D.

In this sense, in a 1D system it is not possible to characterise clear
regions where the currents decay and do not decay, but rather the
crossover must be characterised by how rapidly the current decays for
different initial mean quasi-momenta and inter-particle interaction
strengths. In section~\ref{sec:currentresults} we will investigate this
crossover quantitatively in terms of such decay rates using the iTEBD
algorithm.

\subsection{The quasi-condensate fraction}
\label{sec:quasi-cond-fract}

In section~\ref{sec:currentresults} we show that the decay of the
current results from two-particle scattering processes that broaden
the quasi-momentum distribution. In 2D or 3D decay of the current is
thus also directly linked to a decrease in the condensate fraction in
the experiment, which was used as the key experimental observable in
reference~\cite{mun07}. In 1D, there is no condensate for an infinite
system, however as the system is finite (of the order of 100 occupied
lattice sites), a peak that would correspond to the condensate is
observed in time of flight measurements of the momentum
distribution. In order to make connection to this observable, we will
calculate not only the time dependence of the current $\langle \hat j
\rangle$ below, but also a \textit{quasi-condensate} fraction defined
over a finite portion of the system. This fraction, $\mathcal{C}_R$,
can be defined as the largest eigenvalue of the reduced SPDM
considering only a range of $R$ sites in the system i.e.  $\langle
\hat b_i^\dag \hat b_j \rangle$, with $i, j \in [l,l+R]$ for some
arbitrary site $l$. We typically use $R=100$ below, reflecting the
typical occupied number of lattice sites in experiments.

\section{Numerical Calculations of the current decay}
\label{sec:numprod}

In the remainder of this manuscript, we analyse the decay of a
current, which is created by suddenly imparting a quasi-momentum $ka$
at time $t=0$ to bosons loaded into the lattice in the ground state
configuration (which has zero mean quasi-momentum). This resembles the
experimental situation of reference~\cite{mun07}, where a current is
obtained by means of a moving optical lattice which is accelerated to
a final quasi-momentum $ka$ on a timescale long enough to ensure
adiabaticity with the respect to inter-band transitions. Here we
assume that the acceleration is fast compared with the tunnelling
timescale $1/J$, resulting in a simple translation of the
quasi-momentum distribution, (see section~\ref{sec:initial-current}
and figure~\ref{fig:mdfshift}). This corresponds to an application of
the operator
\begin{equation}\label{eqn:kickop}
  \hat K(ka) 
  \equiv 
  \prod_{l} e^{-i(ka)l\hat n_l}
\end{equation}
to the initial ground state.

To compute both the initial ground state of the system, and the time
dependence of the current once the initial mean quasi-momentum $ka$ is
imparted on the system, we make use of the iTEBD algorithm
\cite{vidal07}. This algorithm is an extension of the TEBD algorithm
\cite{vidal03,vidal04}, and makes possible the near-exact integration
of the Schr\"odinger equation for 1D lattice Hamiltonians in an
infinite, translationally invariant system. These methods have not
only been applied to coherent dynamics in 1D, but also generalised to
dissipative systems \cite{verstraete04, zwolak04, daley09}, and
extensions to a 2D state ansatz have been recently considered
\cite{verstraete08}.

Until now, iTEBD has been primarily used to compute the ground state
of translationally invariant systems, however here this algorithm
becomes crucial in computing time-dependent dynamics
\footnote{Time-dependent dynamics with the iTEBD algorithm have also
  been recently employed to study quantum quenches
  \cite{barmettler09}.}. By computing the decay of the current in an
effectively infinite system, we avoid difficulties arising due to
either open boundary conditions, which can prevent propagation of
moving bosons, and periodic boundary conditions, which restrict the
length over which correlation functions can be computed. In our
calculations we apply the iTEBD algorithm as described in
reference~\cite{vidal07}, but with the modification that the length of
a single block in the translationally invariant system is increased
from two to $2 \pi/ka$. Although a matrix product state representation
can reproduce phase relationships with any period, even with a single
repeated matrix, we find it convenient to increase the block size so
as to represent one full cycle of the phase when we apply the operator
$\hat K(ka)$ to the state (see \ref{app:blockconv} for more details).

Throughout our calculations, we performed convergence tests in the
matrix product state bond-dimension $\chi$, as well as the time step
for integration, and the maximum number of particles allowed per site
in our description of the Hilbert space, $d$. For a more detailed
analysis of how errors enter this method, see \cite{gobert05}.

\section{Analysis of Current decay}
\label{sec:currentresults}

In this section we characterise the current and its decay as a
function of the initial mean quasi-momentum $ka$ and the strength of
interactions $u$, as well as the evolution of the characteristics of
the state as a function of time. We begin in
section~\ref{sec:initial-current} by analysing the initial state of
the system at a time $t=0^+$, after the instantaneous acceleration of
the lattice results in a mean initial quasi-momentum $ka$.  In
section~\ref{sec:evol-char-state} we then describe the dynamical
changes in the quasi-momentum distribution during time
evolution. Finally, we discuss the dependence of the decay on $ka$ and
$u$ and provide a shaded plot of a stability crossover diagram in
section~\ref{subsec:parameterdep}.

\subsection{Initial current}
\label{sec:initial-current}

In this subsection we characterise the system configuration at a time
$t=0^+$ immediately following the initial instantaneous acceleration
(corresponding to an imposition of the initial mean quasi-momentum
$ka$). We consider both the quasi-momentum distribution $n_q$ and the
corresponding current $\langle \hat j \rangle$.

\begin{figure}[htb] 
  \centering
 \includegraphics[width=0.7\textwidth]{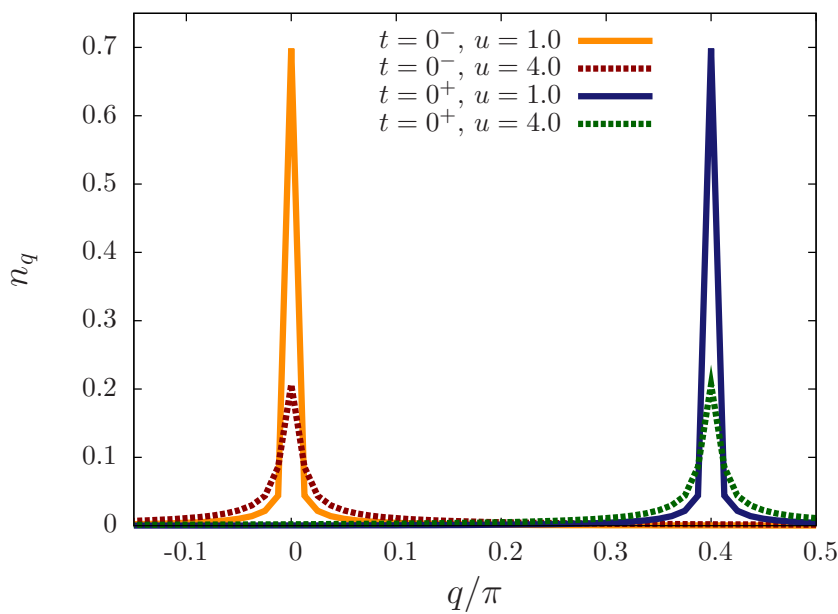}
 \caption{ \label{fig:mdfshift} Normalised quasi-momentum
    distributions before ($t=0^-$) and after ($t=0^+$) a momentum
    $ka=0.4\pi$ is imparted on the particles. The solid lines show the
    distributions for an inter-particle interaction of $u=1$ (SF
    regime), the dotted lines are for distributions with $u=4$ (MI
    regime). The numerical parameters are $\chi=100$, $d=6$.}
\end{figure}

Figure~\ref{fig:mdfshift} shows $n_q$ before and after the
instantaneous application of a momentum shift of $ka=0.4 \pi$ at $t=0$
for a situation where the lattice filling is $\bar n=1$. At $t=0^-$
(that is, before the shift is applied) the system is in a SF or MI
configuration, with $u=1$ and $4$, respectively. At time $t=0^-$ and
for $u=1$, $n_q$ is strongly peaked around $q=0$, as expected for a
superfluid ground state, while it broadens as interactions are
increased, as shown for $u=4$.  The shift $ka$ is applied
instantaneously, and the shape of $n_q$ is unaltered at time
$t=0^+$. This latter distribution is the initial condition for the
subsequent analysis of the system dynamics.

\begin{figure}[htb] 
  \centering
 \includegraphics[width=0.7\textwidth]{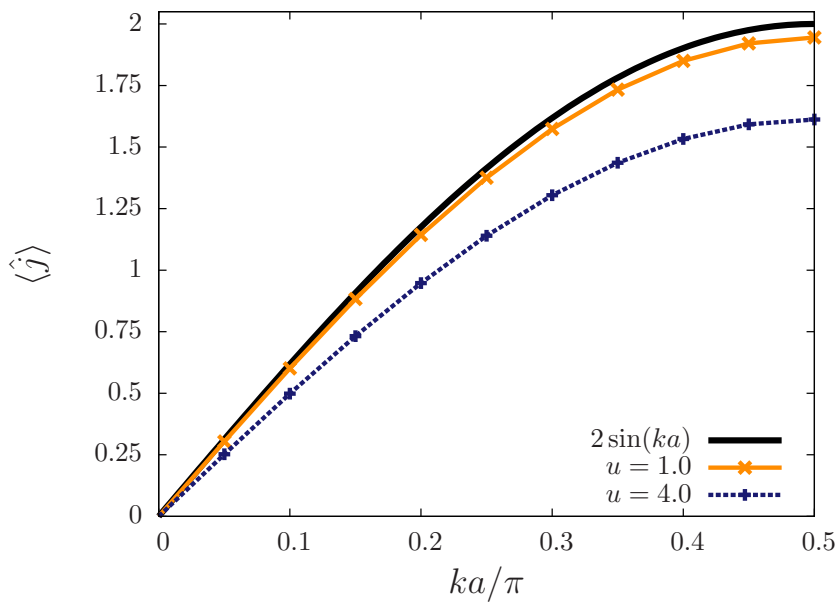}
 \caption{\label{fig:initcurrent} Boson current $\langle \hat j
   \rangle$ in units where $J=1$ after quasi-momenta $0<ka/\pi< 0.5$
   are imparted (time $t=0^+$). In the limit $u\to 0$, $\langle \hat j
   \rangle = 2 J \sin(ka)$ (solid black line). Shown are results for
   interactions $u=1$ and $u=4$ with a filling factor of $\bar n=1$
   particles per site. The numerical parameters are $\chi=100$,
   $d=6$.}
\end{figure}

Figure~\ref{fig:initcurrent} shows the current $\langle \hat j
\rangle$ calculated at $t=0^+$ for a few values of $u$. In the limit
of vanishing interactions, the current $\langle \hat j \rangle $ at
$t=0^+$ is $\langle \hat j \rangle = 2 J \bar n \sin{(ka)}$,
corresponding to all particles occupying the same quasi-momentum
state. In a MI state with integer $\bar n$, this current will be zero
in the limit of infinitely strong interactions, as the quasi-momentum
distribution $n_q$ becomes flat. In figure~\ref{fig:initcurrent} the
case with $u=0$ is plotted as a solid black line. The figure shows
that for finite values of the interaction strength $u$ from
figure~\ref{fig:mdfshift}, $\langle \hat j \rangle$ decreases from the
value $2 J \bar n \sin{(ka)}$ as expected. However, we note that the
current is comparatively large even for the initial condition $u=4$,
corresponding to a MI state at $t=0^-$.  This is related to the fact
that the quasi-momentum distribution $n_q$, whilst broader than in the
superfluid case, is still finite in width (see
figure~\ref{fig:mdfshift}), and far from the flat distribution
corresponding to the limit $u\to \infty$. As $u$ is increased, the
distribution continues to broaden gradually, but there is no strong
qualitative change in the initial quasi-momentum distribution or the
initial current values on entering the Mott Insulator regime.

\subsection{Time evolution of $n_q$}
\label{sec:evol-char-state}

Due to the lack of Galilean invariance in a lattice, a state with a
finite current does not correspond to an eigenstate of the system. We
expect that in the presence of interactions any initial current
$\langle \hat j \rangle $ will eventually decay to a current-free
state.  The dynamics of the current decay corresponds to the
rearrangement of the quasi-momentum distribution by two-atom
scattering processes. The quasi-momentum of two particles is always
conserved in these scattering processes, but because of Umklapp
processes that connect the two edges of the Brillouin zone, the mean
quasi-momentum is significantly changed, and will reach zero in a
final steady state.

\begin{figure}[htb]
  \centering
  \includegraphics[width=0.7\textwidth]{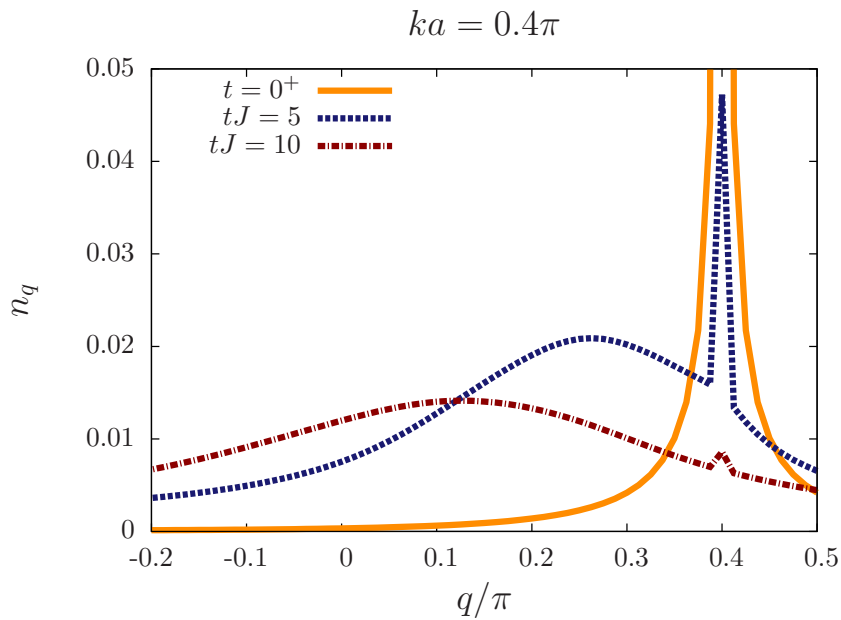}
\caption{ \label{fig:mdfdec} Time evolution of the normalised ground
   state quasi-momentum distribution $n_q$ in the SF regime for $u=1$
   after a momentum $ka=0.4\pi$ is imparted. The particles are
   redistributed into a final state with a broader distribution at
   smaller mean momentum on the timescale $10/J$ (see text). The
   numerical parameters are $\chi=100$, $d=6$.}
\end{figure}

In figure~\ref{fig:mdfdec} we show results for the time evolution of
$n_q$ for a system with the initial conditions $ka=0.4\pi$ and $u=\bar
n= 1$. We observe for increasing time that the width of the momentum
distribution increases and its mean value shifts towards zero.  We
have checked that the off-diagonal elements of the SPDM decay
exponentially in the final state. These results are consistent with
those reported in reference~\cite{rigol05}, where the change in the
redistribution of quasi-momentum in the final state is linked to the
appearance of a finite temperature in the system \cite{rigol08,
  rigol07, kollath07, manmana07}.

We find the behaviour of $n_q$ and SPDM described above to be typical
of all parameter values we have investigated, with the primary
difference for different parameters being in the timescale on which
these processes occur. This includes the Mott Insulator regime, where
the main qualitative difference is in the timescale of the decay of
the current, not in the form in which the quasi-momentum distribution
is redistributed. The dependence of this timescale on the different
system parameters will be investigated below.

\subsection{Stability diagram for the current}
\label{subsec:parameterdep}

\begin{figure}[b]
  \centering
 \includegraphics[width=0.70\textwidth]{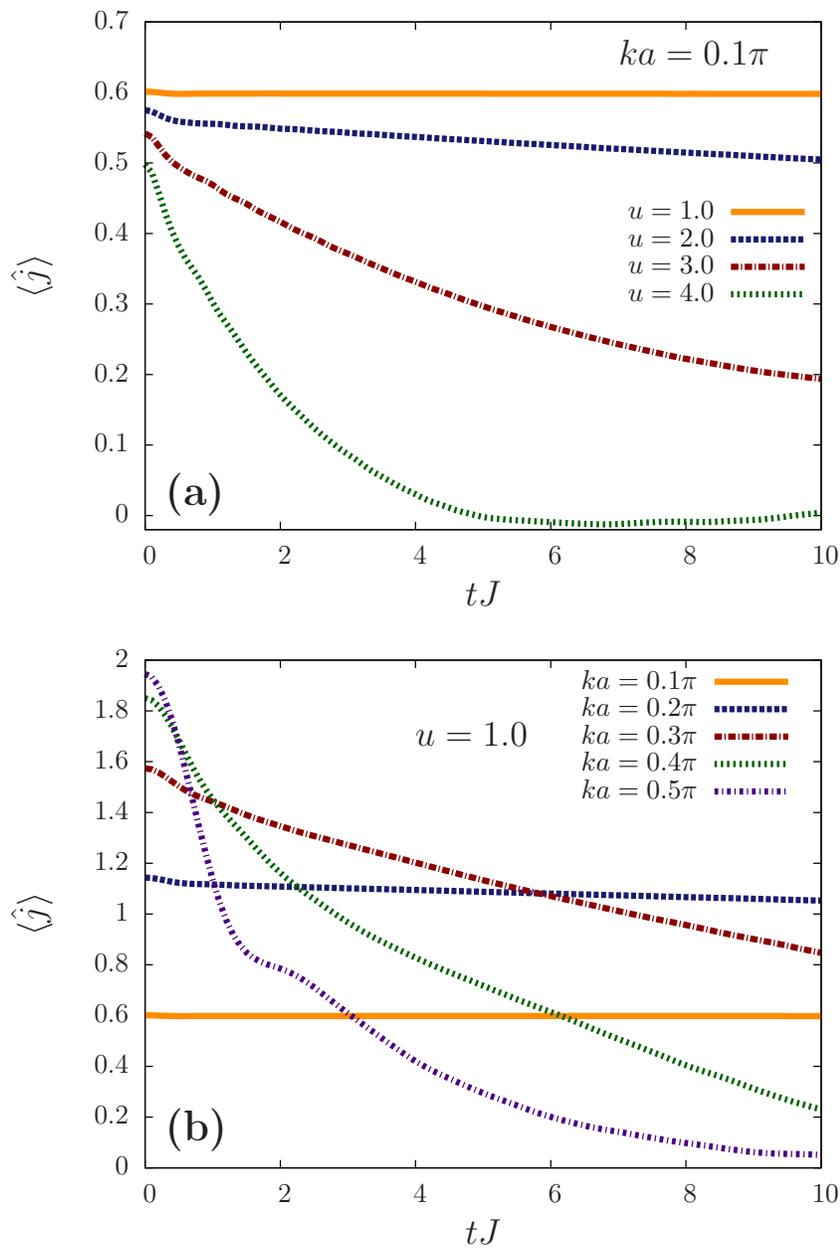}
 \caption{\label{fig:decay} Time evolution of the boson current
    $\langle \hat j \rangle$ for different initial parameters $u$ and
    $ka$. In panel (a) results are compared for increasing values of
    the interaction $u=1,2,3,4$ for a constant small initial mean
    momentum $ka=0.1\pi$. Panel (b) shows results for increasing
    initial momenta $ka=0.1\pi, 0.2\pi, 0.3\pi, 0.4\pi, 0.5\pi$ for
    $u=1$. The numerical parameters are $\chi=100$, $d=6$.}
\end{figure}

We now investigate the decay of the current as a function of the
initial mean quasi-momentum $ka$ and the inter-particle interactions
$u$ for an initial ground state with $\bar n=1$. We focus on the
time-dependent dynamics of $\langle \hat j \rangle $ and also of the
condensate fraction $\mathcal{C}_{100}$ introduced in
section~\ref{sec:quasi-cond-fract}. We extract a stability diagram for
the crossover between parameter regions where currents are stable over
long time periods, and regimes where currents decay
rapidly.

In the following we focus on systems with interaction strengths $0.5
\le u \le 4.0$ and $0.05\pi \le ka \le 0.50\pi$, and we compute the
real time evolution of the boson current and the condensate fraction
$\mathcal{C}_{100}$ for times $0 \le tJ \le 10$. This time interval is
sufficiently short to allow currents to be probed in the experiment
within typical decoherence times, e.g., due to incoherent light
scattering \cite{jaksch05}, and also allows for accurate numerical
results.

\begin{figure}[htb]
  \centering
\includegraphics[width=0.70\textwidth]{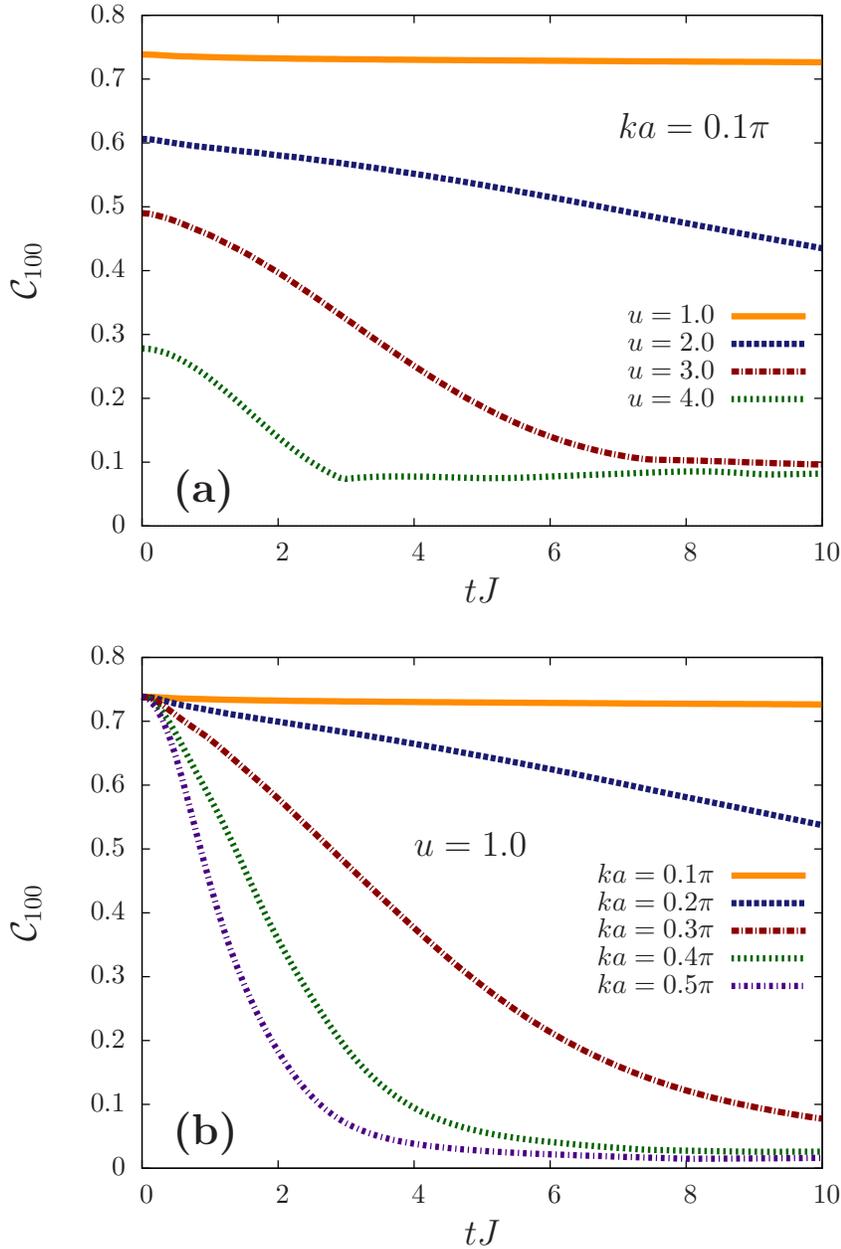}
\caption{ \label{fig:cfdecay} Time evolution of the condensate
   fraction $\mathcal{C}_{100}$ evaluated from the SPDM over a range
   of $100$ sites from the infinite homogeneous system. Qualitatively
   the behaviour is analogous to the boson current in
   figure~\ref{fig:decay}. Panel (a) shows the decay for constant
   initial mean momentum $ka$ and varying interactions $u$, and panel
   (b) for constant $u$ and various $ka$. The numerical parameters are
   $\chi=100$, $d=6$.}
\end{figure}

Figure~\ref{fig:decay}(a) shows the time evolution of $\langle \hat j
\rangle$ for a constant initial momentum $ka=0.1\pi$ and increasing
values of the on-site interaction $u$, and figure~\ref{fig:decay}(b)
shows the decay of $\langle \hat j \rangle$ for a fixed on-site
interaction of $u=1.0$ and various values of $ka$. On the considered
timescale we find that a constant superfluid current (with no decay)
only exists for small values of $ka$ and $u$, for example $ka=0.1 \pi$
with $u=1.0$. For a slightly increased initial quasi-momentum or
on-site interaction a small decay of about $10\%$ on the timescale of
the simulation becomes visible.  Further increasing $ka$ and/or $u$
causes the decay rate to increase rapidly until for $u \gtrsim 3.0$ at
$ka=0.1\pi$ and $ka \gtrsim 0.3\pi$ at $u=0.1$, the current has
completely vanished at time $tJ=10$. Figure~\ref{fig:cfdecay} shows
that the dependence of the time evolution of the condensate fraction
$\mathcal{C}_{100}$ on $ka$ and $u$ is qualitatively very similar to
that of $\langle \hat j \rangle $.

An important result seen in both figure~\ref{fig:decay} and
figure~\ref{fig:cfdecay} is that we do not find a sharp transition point
for $u$ and $ka$ separating parameter regions where the current is
stable or unstable. This lack of a sharp transition is expected for
one-dimensional systems, and is in contrast to the results found in
higher dimensions
\cite{altman05}.

\begin{figure}[htb]
  \centering
  \includegraphics[width=0.70\textwidth]{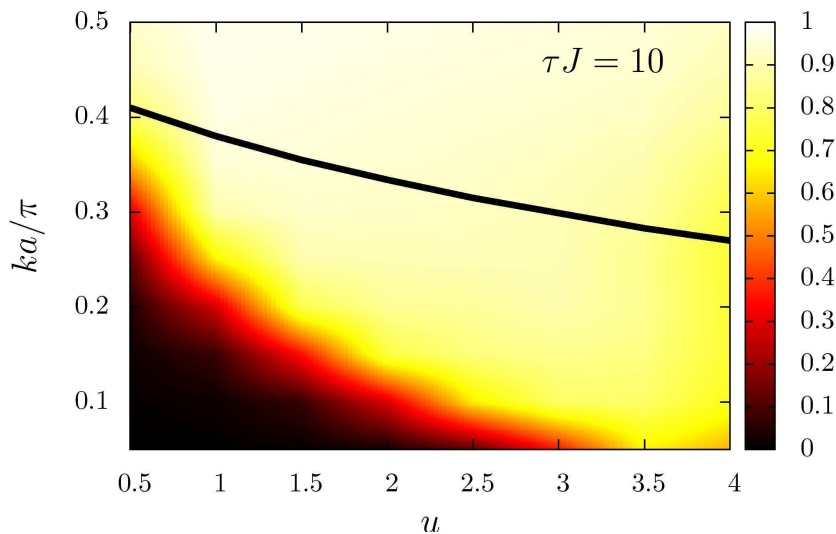}
  \caption{\label{fig:stabcrossover} Shaded plot of the loss of the
   condensate fraction $\Delta \mathcal{C}_{100}$ as a function of $u$
   and $ka$ at a fixed time $\tau J=10$.  Values are shown on a
   $u$-$ka$ grid with spacings of $\Delta u=0.5$ and $\Delta ka =
   0.05\pi$, smoothed by a spline interpolation.  The mean-field
   Gutzwiller prediction from \cite{altman05, polkovnikov05} for the
   transition is drawn as a solid black line. The figure visualises a
   stability crossover diagram (see text).  The numerical parameters
   are $\chi=100$, $d=6$.}
\end{figure}

\begin{figure}[htb]
  \centering
 \includegraphics[width=0.70\textwidth]{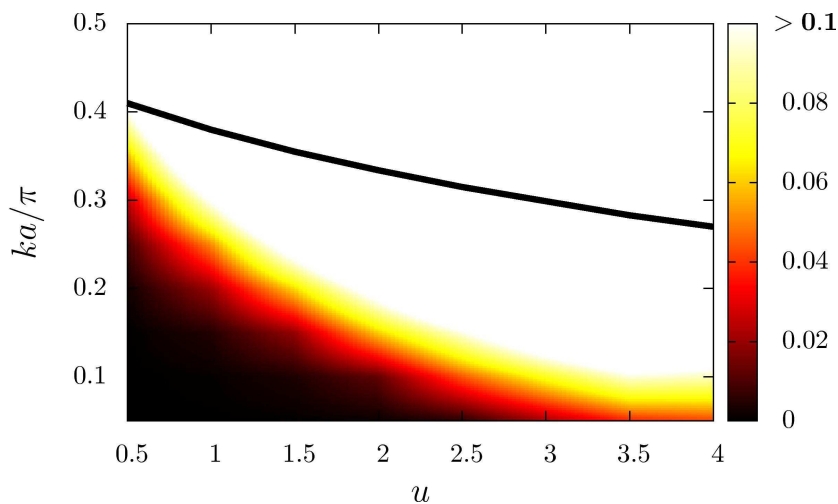}
 \caption{\label{fig:decaygrid} Shaded plot of the condensate
    fraction decay rates $\Gamma_{\mathcal{C}}/J$ as a function of $u$
    and $ka$. The decay rates are extracted from the time evolution of
    the condensate fraction $\mathcal{C}_{100}$ by a linear fitting
    technique (see text for details).  Values are shown on a $u$-$ka$
    grid with spacings of $\Delta u=0.5$ and $\Delta ka = 0.05\pi$,
    smoothed by a spline interpolation.  All decay rates
    $\Gamma_{\mathcal{C}} /J > 0.1$ are marked in white colour. The
    Gutzwiller mean-field prediction is drawn as a solid black
    line. The figure is in qualitative agreement with
    figure~\ref{fig:stabcrossover} and visualises a time-independent
    stability crossover diagram (see text). The numerical parameters
    are $\chi=100$, $d=6$.}
\end{figure}

Figures~\ref{fig:stabcrossover} and \ref{fig:decaygrid} summarise the
dependence of the current decay on $ka$ and $u$, and amount to
stability diagrams for the crossover between parameter regimes where
the current is stable and unstable. In particular,
figure~\ref{fig:stabcrossover} is a shaded plot of the loss of the
condensate fraction $\Delta \mathcal{C}_R(\tau)$, which we define as
\[
\Delta \mathcal{C}_{R}(\tau) \equiv 
| \mathcal{C}_{R}(t=\tau) - \mathcal{C}_{R}(t=0) | / \mathcal{C}_{R}(t=0).
\]
We plot $\Delta \mathcal{C}_{R}(\tau)$ as a function of $ka$ and $u$
at a {\it specific time } $\tau J=10$, as may be measured directly in
an experiment. We chose $R=100$, but found that $\Delta
\mathcal{C}_{R}(\tau) $ is independent of $R$, based on a comparison
of calculations with $R=50$, $100$, and $200$ sites.  We see that
stable currents exist on this timescale only for small values of $u
\lesssim 1.5$ and $ka\lesssim 0.15 \pi$, while no stable current is
present for $u \gtrsim 2.5$ and $ka \gtrsim 0.25\pi$. In the
intermediate region we observe a smooth crossover between those two
regimes. In the classical limit of small on-site interactions, the
stability/instability crossover tends to occur at large values of
$ka\approx 0.5\pi$, which corresponds to the dynamical instability of
reference~\cite{smerzi02}, while for $ka\approx 0$ the instability
sets in close to the value $u_c\approx 3.37$, which corresponds to the
SF-MI transition at zero current \cite{kuhner00}.  As a reference, in
figure~\ref{fig:stabcrossover} we plot the mean-field result of
references~\cite{altman05, polkovnikov05} as a solid black line.  The
latter provides a reasonable indication of the position of the
crossover region for small $u \lesssim 0.5$ only. Outside of this
region, the decay appears typically to be much faster than is expected
from the mean-field estimates.

Note that $\Delta \mathcal{C}_R(\tau)$ discussed above can be directly
observed via interference patterns in momentum distributions
\cite{bloch08}, which can be measured via time-of-flight measurements
in experiments. We can estimate the corresponding experimental
timescales by taking a typical lattice depth of $10 \ E_R$, where
$E_R$ denotes the recoil energy of the atoms $E_R \equiv p_r^2 / 2 m$
with the recoil momentum $p_r \equiv h / \lambda_{l}$. For these
lattice depths, the tunnelling amplitude $J/\hbar$ for Rb atoms is of
the order of $100 \ \mbox{Hz}$. Thus, the timescale $\tau J=10$
corresponds to experimental timescales of the order of $100 \
\mbox{ms}$, which is
within typical coherence times.

Figure~\ref{fig:decaygrid} shows a {\it time-independent} stability
diagram for the current, which we compute from decay rates
$\Gamma_{\mathcal{C}}$ for the condensate fraction
$\mathcal{C}_{100}$.  As was shown in figure~\ref{fig:cfdecay}, the
decay of $\mathcal{C}_{100}$ as a function of $t J$ can be separated
into three regions. Initially, the system shows an initial decay
behaviour on short timescales $t J < 1$, followed by an intermediate
region where the decay is found to be approximately linear in
time. Finally a saturation of the decay process takes place while the
system approaches the zero-current steady state. We therefore extract
the decay rate $\Gamma_{\mathcal{C}}$ by determining the slope of the
decay in the approximately linear intermediate region. Note that in
the short time region where we fit, a linear decay behaviour is also
equivalent to an exponential decay $\mathcal{C}_{100}(t) \propto
\exp(-\Gamma_{\mathcal{C}} t)$.

Figure~\ref{fig:decaygrid}, which is a shaded plot of
$\Gamma_{\mathcal{C}}$, for $\Gamma_{\mathcal{C}}/J \leq 0.1$, shows
results which are qualitatively similar to those of
figure~\ref{fig:stabcrossover}. That is, stable currents are found to
exist only for small values of $ka$ and $u$, and the
stability/instability crossover in general occurs at values of $ka$
and $u$ significantly smaller than predicted by mean-field theory (see
solid black line in the figure).

In the next section we compute decay rates for various lattice
fillings $\bar n \geq 1$ and investigate the scaling with the
interaction strength and filling factor. We find suprisingly that the
scaling is very similar to those computed for a weakly interacting
system with $\bar n \gg 1$ using beyond-mean-field (phase-slip)
calculations.

\section{Scaling of the current decay rates}
\label{sec:phaseslip}

In this section, we study the dependence of the decay rate
$\Gamma_{\mathcal{C}}$ defined in section~\ref{subsec:parameterdep}
and of an analogous time-independent decay rate $\Gamma_j$ for the
current (defined below) on the lattice filling $\bar n$, the
interaction strength $u$, and the quasi-momentum $ka$. We find scaling
laws exhibiting exponential dependence on these
quantities. Interestingly, these scaling laws are similar to those
computed for current decay in a weakly interacting system with large
filling factor $n\gg 1$, which were computed via (beyond-mean-field)
instanton calculations in reference~\cite{polkovnikov05}. This is
especially true in the case $\bar n \geq 3$, or far from the SF-MI
transition with $\bar n = 1$. Naturally, significant deviations occur
between the exact numerical results close to the SF-MI transition and
predictions for a weakly interacting system. Note that here we do not
attempt to verify these instanton predictions within their regime of
validity, which has been done in a recent study by Danshita and
Polkovnikov \cite{danshita09_2}, using particle numbers $\bar n \sim
1000$. Instead, we are interested in exploring the regimes close to
$\bar n = 1$, which correspond to recent experiments with atoms
confined in 3D optical lattices.

In reference~\cite{polkovnikov05}, a decay rate $\Gamma$ for the
current in a system with large filling factor was calculated using
instanton (phase-slip) techniques. This could be shown to be $\Gamma
\propto e^{-S}$, with $S$ the semi-classical action
\begin{equation} \label{eqn:tunnelingaction}
  S\approx 
  7.1\sqrt{\frac{\bar n}{u}}
  \left(
    \frac{\pi}{2} - ka
  \right)^{(5/2)}
.
\end{equation}
Equation~(\ref{eqn:tunnelingaction}) is strictly valid in 1D in the
weakly interacting (Gross-Pitaevskii) limit, $u / \bar n \ll 1$, for
$u \bar n \gg 1$, i.e. in the limit of large filling factors, and
close to the classical instability at $ka\approx \pi/2$. This is very
different from the parameter regimes that are typical when the 1D
system is formed via a 3D optical lattice, as we study here in the
present article. In principle, our filling factors $\bar n \leq 4$,
and acceleration values $ 0.2 \leq ka/\pi \leq 0.35$, do not correpond
to the region of validity of equation~(\ref{eqn:tunnelingaction}).
However, as we will see below, we nonetheless obtain very similar
scalings.
\begin{figure}[htb] 
  \centering
  \includegraphics[width=0.70\textwidth]{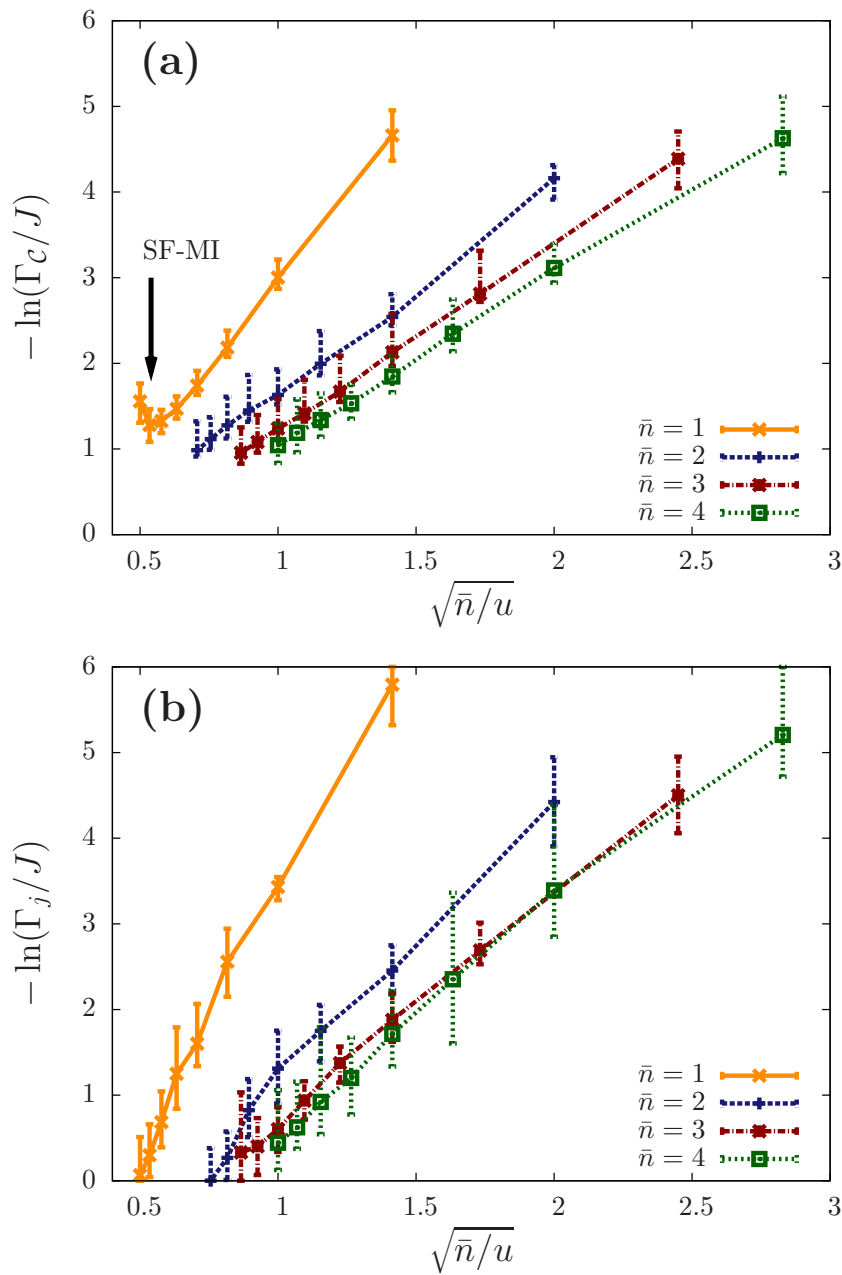}
  \caption{\label{fig:actionscaling} The negative logarithm of the
   decay rate of $\mathcal{C}_{100}$ in panel (a), and of the boson
   current in panel (b) as a function of $\sqrt{\bar n/u}$ for a fixed
   initial mean momentum $ka=\pi/4$. The results are estimated from
   linear decay rate fits. Both panels visualise a similar linear
   dependence (see text). The numerical parameters are $\chi=50$,
   $d=10$.}
\end{figure}

Figures~\ref{fig:actionscaling}(a) and \ref{fig:actionscaling}(b) show
our results for the decay rates $\Gamma_{\mathcal{C}}$ and $\Gamma_j$,
respectively, where $\Gamma_j$ has been extracted from the time
evolution of the boson current $\langle \hat j \rangle$ (see
figure~\ref{fig:decay}) in the same way as for $\Gamma_{\mathcal{C}}$
as described in section~\ref{subsec:parameterdep}. In panels (a) and
(b) we plot $-\ln (\Gamma_{\mathcal{C}}/J)$ and $-\ln (\Gamma_j/J)$
respectively as a function of $\sqrt{\bar n/u}$, for $1 \leq \bar n
\leq 4$ and a fixed $ka=\pi/4$. Both of these plots show an
approximately linear dependence on $\sqrt{{\bar n}/{u}}$, for all
$\bar n$ and $\sqrt{{\bar n}/{u}} \gtrsim 0.6$. In addition, for these
values of $\sqrt{{\bar n}/{u}} $ both panels show that by increasing
$\bar n$ the slope of the lines tend to converge to a fixed
value. This indicates that not only is the decay rate proportional to
$\exp(-\lambda \sqrt{1/u})$, but that the constant $\lambda$ appears
to scale as $\sqrt{\bar n}$ for $\bar n \gtrsim 2$. This is the same
scaling as predicted for the case $\bar n \gg 1$ via instanton
methods. Averaging over the slopes of the results for the large
filling factors $\bar n = 3, 3.5$ and $4$, in order to obtain the
prefactor in $\lambda$, we find values of $\lambda = (2.1\pm
0.5)\sqrt{\bar n}$ and $\lambda = (2.7 \pm 0.6)\sqrt{\bar n}$ for
panels (a) and (b) respectively.  We note that the equality of the
slopes obtained from the current decay and condensate fraction decay
within the fitting error indicates the equivalence of these quantities
as good dynamical observables, for the characterisation of the current
decay both in the numerical simulations and in experiments, for large
enough $\bar n$. For small $\bar n \approx 1$ we observe slightly
different behaviour (see below for more details).

The results above for large $\bar n$ surprisingly show good agreement
with scalings from the instanton calculations for a very different
parameter regime, although we note that the prefactor in $\lambda$ is
different, as the scaling law in equation~(\ref{eqn:tunnelingaction}),
would imply $\lambda= 3.9\sqrt{\bar n}$. The most obvious deviation
between the iTEBD results and equation~(\ref{eqn:tunnelingaction})
takes place for the $\Gamma_{\mathcal{C}}$ results of panel (a) in the
vicinity of the SF-MI transition with $\bar n=1$, which occurs at
$\sqrt{\bar n/u}\approx 0.54$. For $\sqrt{\bar n/ u} \lesssim 0.6$ the
rate $\Gamma_{\mathcal{C}}$ no longer increases as $\sqrt{\bar n/ u}$
decreases. We indicate this parameter region with an arrow in
figure~\ref{fig:actionscaling}(a). This behaviour is not captured by
the decay of the current $\Gamma_j$ in panel (b). This contrast
between the the decay of the condensate fraction and the decay of the
current could reflect a reduced role of the condensate fraction in the
overall system dynamics when we have an initial MI state.  Note that
for $\sqrt{\bar n/u} \lesssim 0.6$ the decay of $\langle \hat j
\rangle$ occurs rapidly on the timescale of our calculation time
steps, which makes computing its decay rate less accurate in this
regime.

\section{Summary and Outlook}
\label{sec:summary}
In summary, the iTEBD method allows us to make quantitative
predictions for the time-dependence of currents for bosons moving in
1D in an optical lattice. In contrast to the behaviour in higher
dimensions, we observe a broad crossover between regions of stable and
unstable currents, with the typical decay rates for the currents being
more rapid than mean-field predictions. We also find surprising
agreement with the scaling of decay rates with interaction strength
and filling fraction calculated for the case of large fillings and
weak interactions via phase-slip methods. These results should be
accessible in current experiments with filling factors near unit
density. As the density increases, we find that the scaling of the
decay rate of the current almost agrees with predictions from
instanton calculations.

These results are strongly related to the decay of currents generated
in a displaced harmonic trap, which has also been the subject of
significant experimental \cite{burger01, cataliotti03, fertig05} and
theoretical \cite{smerzi02, polkovnikov04, gea-banacloche06}
investigation. An understanding of these currents is also an important
starting point for the investigation of more general transport
properties of bosons in 1D. This could include behaviour of these
currents in the presence of an impurity \cite{micheli04,daley05}, or
spatially varying interactions \cite{daley08, tokuno08}.

\ack

We thank D. Jaksch, S. Clark, E. Demler and P. Zoller for helpful
discussions. This work was supported by the Austrian Science
Foundation (FWF) through SFB F40 FOQUS and project I118\_N16
(EuroQUAM\_DQS), the DARPA OLE program, STREP FP7-ICT-2007-C project
NAME-QUAM, and the Austrian Ministry of Science BMWF as part of the
UniInfrastrukturprogramm of the Forschungsplattform Scientific
Computing and of the Centre for Quantum Physics at LFU Innsbruck.

\appendix
\setcounter{section}{0}

\section{Translationally invariant block size and the iTEBD Algorithm}
\label{app:blockconv}

The original iTEBD algorithm as introduced in \cite{vidal07} uses
translationally invariant matrix product states to represent the
quantum state of an infinite homogenous system. In order to conveniently
compute time-evolution, the translationally invariant state is usually represented
by a repeated block of two sites, as described in reference \cite{vidal07}.

\begin{figure}[htb] 
  \centering
  \includegraphics[width=0.70\textwidth]{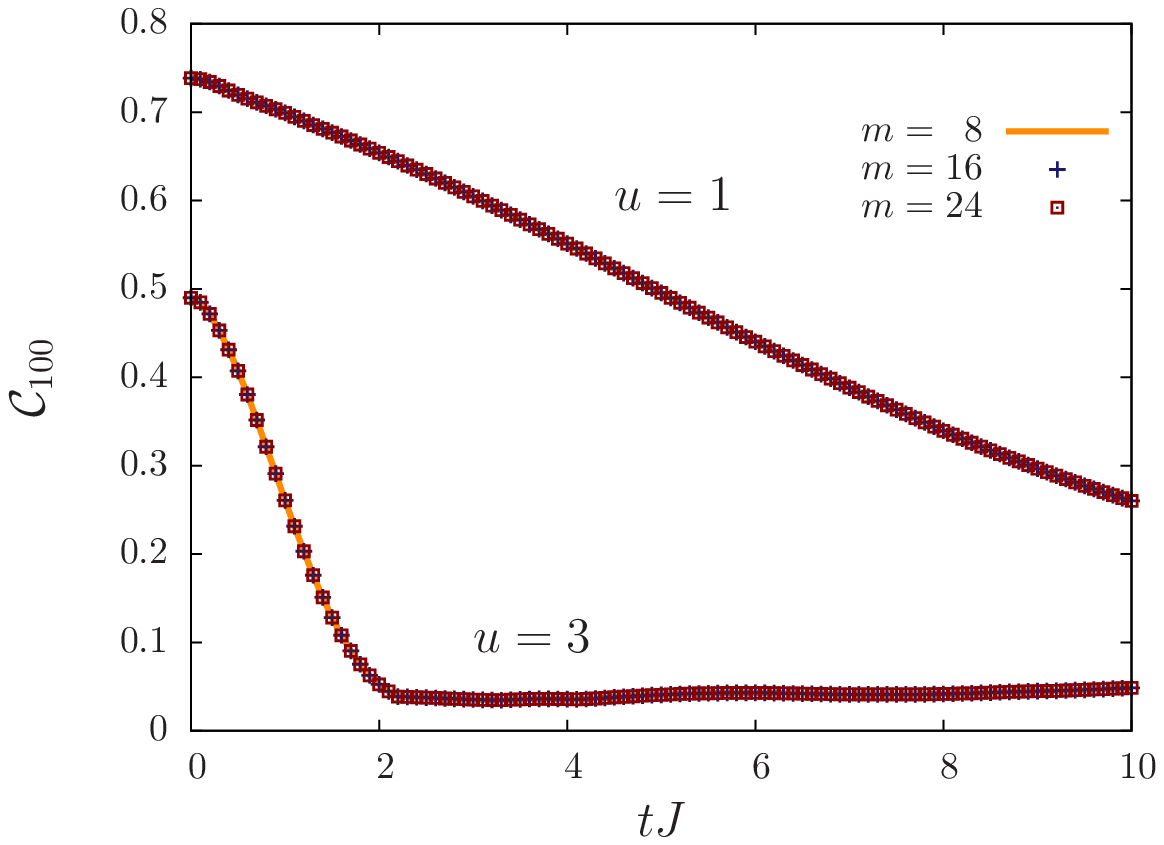}
  \caption{\label{fig:blockconv} The time evolution of
    $\mathcal{C}_{100}$ after an initial quasi-momentum acceleration
    $ka=\pi/4$ for two different on-site interactions $u=1$ and
    $u=3$. No differences are visible when doubling or tripleing the
    block size $m$ within the iTEBD algoritm. Other numerical
    parameters are $\chi=100$, $d=6$.}
\end{figure}

Note that despite the block size of two, excitations with any period
can be represented in this form, because matrix product states
effectively store phase relationships between neighbouring sites
rather than phases corresponding to a single site. Indeed, the use of
a block size of two is a matter of convenience in applying the
algorithm, and if variational methods are applied \cite{verstraete08},
then a block size of one can be sufficient to represent a
translationally invariant state. In the present work, we find it
convenient to extend the block size beyond two to $m$ sites, where for
a particular simulation in which the momentum translation of the
initial state is $ka$, we choose $m=2\pi/ka$. This is convenient in
the application of the operator $\hat K(ka)$, but has no fundamental
effects on the computation of time evolution of the state. In our
case, time evolution is simulated by decomposing the time evolution
operator into $m$ two-site unitary operations, instead of two as in
the original iTEBD algorithm. We expect that this, in case of a large
enough bond dimension $\chi$ still results in an exact state
representation for the infinite system during the whole time
evolution.

To demonstrate the independence of the computation of time evolution
on our choice of $m$, we plot example values for the time evolution of
the condensate fraction $\mathcal{C}_{100}$, computed with increasing
$m$ in figure~\ref{fig:blockconv}. We use an initial acceleration
$ka=\pi/4$, and compare results for block sizes $m=8,16,24$, for two
different values of the interaction strength $u=1, 3$. We see that the
results are independent of the block size, and after testing for
convergence in the parametes $\chi$ and $d$ should represent exact
time-evolved values.

\end{document}